\documentclass{article}
\usepackage{amsfonts}

\usepackage{graphicx}
\usepackage{amsmath}

\newtheorem{theorem}{Theorem}

\begin{document}

\title{Quantum White Noises and The Master Equation for Gaussian Reference States}
\author{John Gough \\
Department of Computing and Mathematics,\\
Nottingham Trent University,\\
Nottingham\\
NG1 4BU\\
United Kingdom}
\date{}
\maketitle

\begin{abstract}
We show that a basic quantum white noise process formally reproduces quantum
stochastic calculus when the appropriate normal / chronological orderings
are prescribed. By normal ordering techniques for integral equations and a
generalization of the Araki-Woods representation, we derive the master and
random Heisenberg equations for an arbitrary Gaussian state: this includes
thermal and squeezed states.
\end{abstract}

\bigskip

\section{Quantum White Noises}

It is possible to develop a formal theory of quantum white noises which
nevertheless provides a powerful insight into quantum stochastic processes.
We present the ``bare bones'' of the theory: the hope is that the structure
will be more apparent without the mathematical gore and viscera.

Essentially, we need a Hilbert space $\mathcal{H}_{0}$\ to describe a system
of interest. We postulate a family of pseudo-operators $\left\{
a_{t}^{+}:t\geq 0\right\} $ called creation noises; a formally adjoint
family $\left\{ a_{t}^{-}:t\geq 0\right\} $ called annihilation noises and a
vector $\Psi $ called the vacuum vector. The noises describe the environment
and are assumed to act trivially on the observables of $\mathcal{H}_{0}$.

The first structural relations we need is 
\begin{equation}
\begin{tabular}{|l|}
\hline
\\ 
$a_{t}^{-}\Psi =0$ \\ 
\\ \hline
\end{tabular}
\tag{QWN1}
\end{equation}
which implies that the annihilator noises annihilate the vacuum. The second
relations are given by the commutation relations

\begin{equation}
\begin{tabular}{|l|}
\hline
\\ 
$\left[ a_{t}^{-},a_{s}^{+}\right] =\kappa \delta _{+}\left( t-s\right)
+\kappa ^{\ast }\delta _{-}\left( t-s\right) $ \\ 
\\ \hline
\end{tabular}
\tag{QWN2}
\end{equation}
Otherwise the creation noises all commute amongst themselves, as do the
annihilation noises. Here $\kappa =\frac{1}{2}\gamma +i\sigma $ is a complex
number with $\gamma >0$. We have introduced the functional kernels $\delta
_{\pm }$ having the action 
\begin{equation}
\int_{-\infty }^{\infty }f\left( s\right) \delta _{\pm }\left( t-s\right)
:=f\left( t^{\pm }\right)
\end{equation}
for any Riemann integrable function $f$.

If the right-hand side of (QWN2) was just $\gamma \delta \left( t-s\right) ,$
then we would have sufficient instructions to deal with integrals of the
noises wrt. Schwartz functions; however introducing the $\delta _{\pm }$%
-functions we can formally consider integrals wrt. piecewise Schwartz
functions as we now have a rule for what to do at discontinuities. In
particular, we can consider integrals over simplices $\left\{ t>t_{1}>\dots
>t_{n}>0\right\} .$ The objective is to use the commutation rule (QWN2) to
convert integral expressions involving the postulated noises $a_{t}^{\pm }$
to equivalent normal ordered expressions, so that we only encounter the
integrals of the $\delta _{\pm }$-functions against Riemann integrable
functions.

We remark that the commutation relations (QWN2) arose from considerations of
Markovian limits of field operators $a_{t}^{\pm }\left( \lambda \right) $ in
the Heisenberg picture satisfying relations of the type $\left[
a_{t}^{-}\left( \lambda \right) ,a_{s}^{+}\left( \lambda \right) \right]
=K_{t-s}\left( \lambda \right) \,\theta \left( t-s\right) +K_{t-s}\left(
\lambda \right) ^{\ast }\,\theta \left( s-t\right) $ where $\theta $ is the
Heaviside function and the right-hand side is a Feynman propagator.

\bigskip

\subsection{Quantum Stochastic Calculus}

\subsubsection{Fundamental Stochastic Processes}

For real square-integrable functions $f=f\left( t\right) $, we define the
following four fields 
\begin{equation}
A^{ij}\left( f\right) :=\int_{0}^{\infty }\left[ a_{s}^{+}\right] ^{i}\left[
a_{s}^{-}\right] ^{j}\,f\left( s\right) \,ds,\qquad i,j\in \left\{
0,1\right\}
\end{equation}
where on the right-hand side the superscript denotes a power, that is $\left[
a\right] ^{0}=1$, $\left[ a\right] ^{1}=a$.

With $1_{\left[ 0,t\right] }$ denoting the characteristic function of the
time interval $\left[ 0,t\right] $, we define the four fundamental processes 
\begin{equation}
A_{t}^{ij}:=A^{ij}\left( 1_{\left[ 0,t\right] }\right) =\int_{0}^{t}\left[
a_{s}^{+}\right] ^{i}\left[ a_{s}^{-}\right] ^{j}\,ds,\qquad i,j\in \left\{
0,1\right\} ;
\end{equation}

\subsubsection{Testing Vectors}

We denote by $\mathcal{F}^{\left( n\right) }$ the Hilbert space spanned by
the symmterized vectors $f_{1}\hat{\otimes}\dots \hat{\otimes}f_{n}:=$ $\
A^{10}\left( f_{1}\right) \dots A^{10}\left( f_{n}\right) \,\Psi $. The
Hilbert spaces $\mathcal{F}^{\left( n\right) }$\ are then naturally
orthogonal for different $n$ and their direct sum is the (Bose) Fock space $%
\mathcal{F}=\oplus _{n=0}^{\infty }\mathcal{F}^{\left( n\right) }$. The
exponential vector with test function $f\in L^{2}\left( \mathbb{R}%
^{+}\right) $ is defined to be 
\begin{equation}
\varepsilon \left( f\right) :=\sum_{n=0}^{\infty }\frac{1}{n!}\hat{\otimes}%
^{n}f.
\end{equation}
Note that $\left\langle \varepsilon \left( f\right) |\varepsilon \left(
g\right) \right\rangle =\exp \gamma \left\langle f|g\right\rangle $ and that 
$\Psi \equiv \varepsilon \left( 0\right) $.

For a subset $\mathcal{T}$ of $L^{2}\left( \mathbb{R}^{+}\right) ,$ the
subset of Fock space generated by the elements of $\mathcal{T}$ is $\mathfrak{E}%
\left( \mathcal{T}\right) =\left\{ \varepsilon \left( f\right) :f\in 
\mathcal{T}\right\} $. In general, we shall denote the Fock space over a
one-particle Hilbert space $\mathfrak{h}$\ by $\Gamma \left( \mathfrak{h}\right)
=\oplus _{n=0}^{\infty }\left( \hat{\otimes}^{n}\mathfrak{h}\right) ;$ thus $%
\mathcal{F}\equiv \Gamma \left( L^{2}\left( \mathbb{R}^{+}\right) \right) $.

\subsubsection{Quantum Stochastic Processes}

Let $\mathcal{H}$\ be a fixed Hilbert space, we wish to consider operators
on the tensor product $\mathcal{H}\otimes \mathcal{F}$. A family of
operators $\left( X_{t}\right) _{t\geq 0}$ defined on a common domain $%
D\otimes \mathfrak{E}\left( \mathcal{T}\right) $ can be understood as a mapping
from $:D\times D\times \mathcal{T}\times \mathcal{T}\times \mathbb{R}%
^{+}\rightarrow \mathbb{C}$ :$\left( \phi ,\psi ,f,g,t\right) \mapsto
\left\langle \phi \otimes \varepsilon \left( f\right) |X_{t}\,\psi \otimes
\varepsilon \left( g\right) \right\rangle $.

\subsubsection{Quantum Stochastic Integrals}

Let $\left( X_{t}^{ij}\right) _{t\geq 0}$ be four adapted processes. The
quantum stochastic integral having these processes as integrands is 
\begin{equation}
X_{t}=\int_{0}^{t}ds\,\left[ a_{s}^{+}\right] ^{i}X_{s}^{ij}\left[ a_{s}^{-}%
\right] ^{j}
\end{equation}
where we use the Einstein convention that repeated indices are summed (over
0,1). We shall use the differential notation $dX_{t}=\left[ a_{t}^{+}\right]
^{i}X_{t}^{ij}\left[ a_{t}^{-}\right] ^{j}\,dt$ or even $\frac{dX_{t}}{dt}=%
\left[ a_{t}^{+}\right] ^{i}X_{t}^{ij}\left[ a_{t}^{-}\right] ^{j}$.

The key feature is that the noises appear in normal ordered form in the
differentials. Suppose that the $\left( X_{t}^{ij}\right) _{t\geq 0}$ are
defined on domain $D\otimes \mathfrak{E}\left( \mathcal{R}\right) $, then it
follows that

\begin{equation*}
\left\langle \phi \otimes \varepsilon \left( f\right) |\frac{dX_{t}}{dt}%
\,\psi \otimes \varepsilon \left( g\right) \right\rangle \equiv \left[
f^{\ast }\left( t\right) \right] ^{i}\left\langle \phi \otimes \varepsilon
\left( f\right) |X_{t}^{ij}\,\psi \otimes \varepsilon \left( g\right)
\right\rangle \left[ g\left( t\right) \right] ^{j}
\end{equation*}
for all $\phi ,\psi \in D$ and $f,g\in \mathcal{R}$. (The equivalence can be
understood here as being almost everywhere.)

\subsubsection{Quantum It\^{o}'s Formula}

Let $X_{t}$ and $Y_{t}$ be quantum stochastic integrals, then the product $%
X_{t}Y_{t}$ may be brought to normal order using (QWN2). In differential
terms we may write this as 
\begin{eqnarray}
d\left( X_{t}Y_{t}\right) &=&\left( dX_{t}\right) Y_{t}+X_{t}\left(
dY_{t}\right)  \notag \\
&=&\left( \hat{d}X_{t}\right) Y_{t}+X_{t}\left( \hat{d}Y_{t}\right) +\left( 
\hat{d}X_{t}\right) \left( \hat{d}Y_{t}\right)
\end{eqnarray}
where the It\^{o} differentials are defined as $\left( \hat{d}X_{t}\right)
Y_{t}:=\left[ a_{t}^{+}\right] ^{i}X_{t}^{ij}\,\left( Y_{t}\right) \,\left[
a_{t}^{-}\right] ^{j}\;dt;$ $X_{t}\left( \hat{d}Y_{t}\right) :=\left[
a_{t}^{+}\right] ^{k}\left( X_{t}\right) \,Y_{t}^{kl}\,\left[ a_{t}^{-}%
\right] ^{l}\;dt;$ $\left( \hat{d}X_{t}\right) \left( \hat{d}Y_{t}\right) :=%
\left[ a_{t}^{+}\right] ^{i}X_{t}^{i1}Y_{t}^{1l}\left[ a_{t}^{-}\right]
^{l}\;dt$.

The quantum It\^{o} ``table'' corresponds to following relation for the
fundamental processes: 
\begin{equation}
\left( \hat{d}A_{t}^{i1}\right) \left( \hat{d}A_{t}^{1j}\right) =\gamma
\left( \hat{d}A_{t}^{ij}\right) .
\end{equation}

\subsubsection{Quantum Stochastic Differential Equations}

The differential equation $\frac{dX_{t}}{dt}=\left[ a_{t}^{+}\right]
^{i}X_{t}^{ij}\left[ a_{t}^{-}\right] ^{j}$ with initial condition $%
X_{0}=x_{0}$ (a bounded operator in $\mathcal{H}_{0}$) corresponds to the
differential equation system 
\begin{equation*}
\left\langle \phi \otimes \varepsilon \left( f\right) |\frac{dX_{t}}{dt}%
\,\psi \otimes \varepsilon \left( g\right) \right\rangle \equiv \left[
f^{\ast }\left( t\right) \right] ^{i}\left\langle \phi \otimes \varepsilon
\left( f\right) |X_{t}^{ij}\,\psi \otimes \varepsilon \left( g\right)
\right\rangle \left[ g\left( t\right) \right] ^{j}
\end{equation*}
and in general one can show the existence and uniqueness of solution as a
family of operators on $\mathcal{H}_{0}\otimes \mathcal{F}$. The solution
can be written as $X_{t}=x_{0}+\int_{0}^{t}ds\,\left[ a_{s}^{+}\right]
^{i}X_{s}^{ij}\left[ a_{s}^{-}\right] ^{j}$. In Hudson-Parthasarathy
notation this is written as $X_{t}=x_{0}+\int_{0}^{t}X_{s}^{ij}\otimes \hat{d%
}A_{t}^{ij}$, where the tensor product sign indicates the continuous tensor
product decomposition $\mathcal{F}=\Gamma \left( L^{2}(0,t]\right) \otimes
\Gamma \left( L^{2}(t,\infty )\right) $.

\subsection{Quantum Stochastic Evolutions}

A quantum stochastic evolution is a family $\left( J_{t}\right) _{t\geq 0}$
mapping from the bounded observables on $\mathcal{H}_{0}$ to the bounded
observables on $\mathcal{H}_{0}\otimes \mathcal{F}$. We are particularly
interested in those taking the form 
\begin{equation*}
J_{t}\left( X\right) \equiv U_{t}^{\dagger }XU_{t}
\end{equation*}
where $U_{t}$ is a unitary, adapted process satisfying some linear qsde (a
stochastic Schr\"{o}dinger equation). In the rest of this section we
establish a Wick's theorem for working with such processes.

\subsubsection{Normal-Ordered QSDE}

Let $V_{t}$ be the solution to the qsde 
\begin{equation}
\frac{dV_{t}}{dt}=L_{ij}\left[ a_{t}^{+}\right] ^{i}V_{t}\left[ a_{t}^{-}%
\right] ^{j},\qquad V_{0}=1;
\end{equation}
where the $L_{ij}$ are bounded operators in $\mathcal{H}_{0}$.\ The
associated integral equation is $V_{t}=1+\int_{0}^{t}dt_{1}\,L_{ij}\left[
a_{t_{1}}^{+}\right] ^{i}V_{t_{1}}\left[ a_{t_{1}}^{-}\right] ^{j}$ which
can be iterated to give the formal series 
\begin{eqnarray}
V_{t} &=&1+\sum_{n=1}^{\infty }\int_{t>t_{1}>\dots t_{n}>0}dt_{1}\dots
dt_{n}\,L_{i_{1}j_{1}}\cdots L_{i_{n}j_{n}}  \notag \\
&&\times \left[ a_{t_{1}}^{+}\right] ^{i_{1}}\cdots \left[ a_{t_{n}}^{+}%
\right] ^{i_{n}}\left[ a_{t_{n}}^{-}\right] ^{j_{n}}\cdots \left[
a_{t_{1}}^{-}\right] ^{i_{1}} \\
&=&\mathbf{\vec{N}}\exp \left\{ \int_{0}^{t}ds\,L_{ij}\left[ a_{s}^{+}\right]
^{i}\left[ a_{s}^{-}\right] ^{j}\right\} ,
\end{eqnarray}
where $\mathbf{\vec{N}}$\ is the normal ordering operation for the noise
symbols $a_{t}^{\pm }$.

Necessary and sufficient conditions for unitary of $V_{t}$ are that 
\begin{equation}
L_{ij}+L_{ji}^{\dagger }+\gamma L_{1i}^{\dagger }L_{1j}=0.
\end{equation}
(Necessity is immediate from the isometry condition $d\left( U_{t}^{\dagger
}U_{t}\right) =U_{t}^{\dagger }\hat{d}\left( U_{t}\right) +\hat{d}\left(
U_{t}^{\dagger }\right) U_{t}+\hat{d}\left( U_{t}^{\dagger }\right) \hat{d}%
\left( U_{t}\right) =U_{t}^{\dagger }\left( L_{ij}+L_{ji}^{\dagger }+\gamma
L_{1i}^{\dagger }L_{1j}\right) U_{t}\otimes \hat{d}A_{t}^{ij}=0$, but
suffices to establish co-isometry $d\left( U_{t}U_{t}^{\dagger }\right) =0$%
.) Equivalently, we can require that 
\begin{equation*}
\text{%
\begin{tabular}{ll}
$L_{11}=\frac{1}{\gamma }\left( W-1\right) ,$ & $L_{10}=L,$ \\ 
$L_{01}=-L^{\dagger }W,$ & $L_{00}=-\frac{1}{2}\gamma L^{\dagger }L-iH;$%
\end{tabular}
}
\end{equation*}
where $W$ is unitary, $H$ is self-adjoint and $L$ is arbitrary.

\subsubsection{Time-Ordered QSDE}

Let $U_{t}$ be the solution to the qsde 
\begin{equation}
\frac{dU_{t}}{dt}=iE_{ij}\left[ a_{t}^{+}\right] ^{i}\left[ a_{t}^{-}\right]
^{j}U_{t},\qquad U_{0}=1;
\end{equation}
where the $E_{ij}$ are bounded operators in $\mathcal{H}_{0}$.\ Here we
naturally interpret $\Upsilon _{t}=E_{ij}\left[ a_{t}^{+}\right] ^{i}\left[
a_{t}^{-}\right] ^{j}$ as a stochastic Hamiltonian. (For $\Upsilon _{t}$ to
be Hermitian we would need $E_{11}$ and $E_{00}$ to be self-adjoint while $%
E_{10}^{\dagger }=E_{01}$.)

Iterating the associated integral equation leads to 
\begin{eqnarray}
U_{t} &=&1+\sum_{n=1}^{\infty }\left( -i\right) ^{n}\int_{t>t_{1}>\dots
t_{n}>0}dt_{1}\dots dt_{n}\,E_{i_{1}j_{1}}\cdots E_{i_{n}j_{n}}  \notag \\
&&\times \left[ a_{t_{1}}^{+}\right] ^{i_{1}}\left[ a_{t_{1}}^{-}\right]
^{i_{1}}\cdots \left[ a_{t_{n}}^{+}\right] ^{i_{n}}\left[ a_{t_{n}}^{-}%
\right] ^{j_{n}} \\
&=&\mathbf{\vec{T}}\exp \left\{ -i\int_{0}^{t}ds\,E_{ij}\left[ a_{s}^{+}%
\right] ^{i}\left[ a_{s}^{-}\right] ^{j}\right\} ,
\end{eqnarray}
where $\mathbf{\vec{T}}$\ is the time ordering operation for the noise
symbols $a_{t}^{\pm }$.

\subsubsection{Conversion From Time-Ordered to Normal-Ordered Forms}

Using the commutation relations (QWN2) we can put the time-ordered
expressions in (12) to normal order. The most efficient way of doing this is
as follows: 
\begin{eqnarray*}
\left[ a_{t}^{-},U_{t}\right] &=&\left[ a_{t}^{-},1-i\int_{0}^{t}ds\,E_{ij}%
\left[ a_{s}^{+}\right] ^{i}\left[ a_{s}^{-}\right] ^{j}U_{s}\right] \\
&=&-i\int_{0}^{t}ds\,E_{1j}\left[ a_{t}^{-\prime }a_{s}^{+}\right] \,\left[
a_{s}^{-}\right] ^{j}U_{s} \\
&=&-i\kappa E_{1j}\,\left[ a_{t}^{-}\right] ^{j}U_{t}
\end{eqnarray*}
or $a_{t}^{-}U_{t}-U_{t}a_{t}^{-}=-i\kappa E_{11}a_{t}^{-}U_{t}-i\kappa
E_{10}U_{t}$. This implies the rewriting rule 
\begin{equation}
a_{t}^{-}U_{t}=\left( 1+i\kappa E_{11}\right) ^{-1}\left\{
U_{t}a_{t}^{-}-i\kappa E_{10}U_{t}\right\} .
\end{equation}
Thus 
\begin{eqnarray*}
iE_{ij}\left[ a_{t}^{+}\right] ^{i}\left[ a_{t}^{-}\right]
^{j}U_{t} &=& iE_{i0}\left[ a_{t}^{+}\right] ^{i}U_{t} \nonumber \\
&& +iE_{i1}\left[ a_{t}^{+}%
\right] ^{i}\left( 1+i\kappa E_{11}\right) ^{-1}\left\{
U_{t}a_{t}^{-}-i\kappa E_{10}U_{t}\right\} . 
\end{eqnarray*}
From this we deduce the
following result:

\begin{theorem}
Time-ordered and normal-ordered forms are related as 
\begin{equation}
\mathbf{\vec{T}}\exp \left\{ -i\int_{0}^{t}ds\,E_{ij}\left[ a_{s}^{+}\right]
^{i}\left[ a_{s}^{-}\right] ^{j}\right\} \equiv \mathbf{\vec{N}}\exp \left\{
\int_{0}^{t}ds\,L_{ij}\left[ a_{s}^{+}\right] ^{i}\left[ a_{s}^{-}\right]
^{j}\right\}
\end{equation}
where 
\begin{equation}
\begin{tabular}{ll}
$L_{11}=-iE_{11}\left( 1+i\kappa E_{11}\right) ^{-1},$ & $L_{10}=-i\left(
1+i\kappa E_{11}\right) ^{-1}E_{10},$ \\ 
&  \\ 
$L_{01}=-iE_{01}\left( 1+i\kappa E_{11}\right) ^{-1},$ & $%
L_{00}=-iE_{00}+\kappa E_{01}\left( 1+i\kappa E_{11}\right) ^{-1}E_{10}.$%
\end{tabular}
\end{equation}
\end{theorem}

\section{Gaussian States}

The above construction requires the existence of a vacuum state $\Psi $; by
application of creation fields we should be able to reconstruct a
Hilbert-Fock space for which $\Psi $ is cyclic. But what about non-vacuum
states? We describe now the trick we shall use in order to consider more
general states for the simple case of one bosonic degree of freedom.

Let $a,a^{\dagger }$ satisfy the commutation relations $\left[ a,a^{\dagger }%
\right] =1$. A state $\left\langle \;\right\rangle $\ is said to be Gaussian
or quasi-free if we have 
\begin{equation}
\left\langle \exp \left\{ iz^{\ast }a+iza^{\dagger }\right\} \right\rangle
=\exp \left\{ \frac{1}{2}\left(2 n+1\right) zz^{\ast }+m^{\ast
}z^{2}+mz^{\ast 2}+iz^{\ast }\alpha +iz\alpha ^{\ast }\right\} ;
\end{equation}
in particular, $\left\langle a\right\rangle =\alpha $, $\left\langle
aa^{\dagger }\right\rangle =n+1$, $\left\langle aa\right\rangle =m$. From
the observation that $\left\langle \left( a+\lambda a^{\dagger }\right)
^{\dagger }\left( a+\lambda a^{\dagger }\right) \right\rangle \geq 0$ it
follows that the restriction $|m|^{2}\leq n\left( n+1\right) $ must apply.

Now suppose that $a_{1},a_{1}^{\dagger }$ and $a_{2},a_{2}^{\dagger }$\ are
commuting pairs of Bose variables and let 
\begin{equation}
a=xa_{1}+ya_{2}^{\dagger }+za_{2}+\alpha
\end{equation}
where $x,y,z,\alpha $ are complex numbers. The commutation relations are
maintained if $|x|^{2}-C|y|^{2}+|z|^{2}=1$. Taking the vacuum state for both
variables $a_{i},a_{i}^{\dagger }$ $\left( i=1,2\right) $ then we can
reconstruct the state $\left\langle \;\right\rangle $\ if $%
|x|^{2}+|z|^{2}=n+1$ and $yz=m$. That is 
\begin{equation}
x=\sqrt{n+1-\frac{|m|^{2}}{n}},\;y=\sqrt{n},\;z=\frac{m}{\sqrt{n}}.
\end{equation}

\subsection{Generalized Araki-Woods Construction}

Let2 $\mathfrak{h}$\ be a fixed one-particle Hilbert space and let $j$ be a
anti-linear conjugation on $\mathfrak{h}$, that is $\left\langle j\phi |j\psi
\right\rangle _{\mathfrak{h}}=\left\langle \psi |\phi \right\rangle _{\mathfrak{h}}$%
\ for all $\phi ,\psi \in \mathfrak{h}$. Denote by $A\left( \phi \right) $ the
annihilator on $\Gamma \left( \mathfrak{h}\right) $ with test function $\phi $
(that is, $A\left( \phi \right) \varepsilon \left( \psi \right)
=\left\langle \phi |\psi \right\rangle _{\mathfrak{h}}\varepsilon \left( \psi
\right) $ ). A state $\left\langle \;\right\rangle $ on $\Gamma \left( \mathfrak{%
h}\right) $ is said to be (mean-zero) Gaussian / quasi-free if there is a
positive operator $N\geq 0$; an operator $M$ with $|M|^{2}\leq N\left(
N+1\right) $ and $\left[ N,M\right] =0$;\ and a fixed anti-linear
conjugation $j$ such that 
\begin{eqnarray}
\left\langle \exp \left\{ iA\left( \phi \right) +iA^{\dagger }\left( \phi
\right) \right\} \right\rangle &=&\exp \bigg\{ -\frac{1}{2}\left\langle \phi
|\left( 2N+1\right) \phi \right\rangle _{\mathfrak{h}} \nonumber \\
&& \qquad 
-\frac{1}{2}\left\langle
M\phi |j\phi \right\rangle _{\mathfrak{h}}-\frac{1}{2}\left\langle j\phi |M\phi
\right\rangle _{\mathfrak{h}} \bigg\} ,
\end{eqnarray}
for all $\phi \in \mathfrak{h}$. In particular, we have the expectations 
\begin{eqnarray*}
\left\langle A\left( \phi \right) A^{\dagger }\left( \psi \right)
\right\rangle &=&\left\langle \phi |N\psi \right\rangle _{\mathfrak{h}} \\
\left\langle A\left( \phi \right) A\left( \psi \right) \right\rangle
&=&\left\langle M\phi |j\psi \right\rangle _{\mathfrak{h}} \\
\left\langle A^{\dagger }\left( \phi \right) A^{\dagger }\left( \psi \right)
\right\rangle &=&\left\langle j\phi |M\psi \right\rangle _{\mathfrak{h}}
\end{eqnarray*}

The state is said to be gauge-invariant when $M=0$. The case where $N=\left(
1-e^{-\beta H}\right) ^{-1}$ and $M=0$ yields the familiar thermal state at
inverse temperature $\beta $ for the non-interacting Bose gas with second
quantization of $H$ as Hamiltonian. The vacuum state is, of course $N=0,M=0$.

The standard procedure for treating thermal states of the interacting Bose
gas is to represent the canonical commutations (CCR) algebra on the tensor
product $\Gamma \left( \mathfrak{h}\right) \otimes \Gamma \left( \mathfrak{h}\right) 
$ and realize state as a double Fock vacuum state. This approach generalizes
to the problem at hand. For clarity we label the Fock spaces with subscripts
1 and 2 and consider the morphism from the CCR algebra over $\Gamma \left( 
\mathfrak{h}\right) $ to that over $\Gamma \left( \mathfrak{h}\right) _{1}\otimes
\Gamma \left( \mathfrak{h}\right) _{2}$ induced by 
\begin{equation}
A\left( \phi \right) \mapsto A_{1}\left( X\phi \right) \otimes
1_{2}+1_{1}\otimes A_{2}^{\dagger }\left( jY\phi \right) +1_{1}\otimes
A_{2}\left( Z\psi \right)
\end{equation}
where

\begin{equation}
X=\sqrt{N+1-\frac{|M|^{2}}{N}},\;Y=\sqrt{N},\;Z=\frac{M}{\sqrt{N}}.
\end{equation}
Here we write $A_{i}\left( \psi \right) $ for annihilators on $\Gamma \left( 
\mathfrak{h}\right) _{i}$ $\left( i=1,2\right) $.

\section{Gaussian Noise}

Now let $a_{1}^{\pm }\left( t\right) $ and $a_{2}^{\pm }\left( t\right) $ be
independent (commuting) copies of quantum white noises with respective vacua 
$\Psi _{1}$ and $\Psi _{2}$. Let $x,y,x$ be as in (18) and let $\alpha $ be
arbitrary complex. We consider the quantum white noise(s) defined by

\begin{equation}
a_{t}^{-}:=xa_{1}^{-}\left( t\right) +ya_{2}^{+}\left( t\right)
+za_{2}^{-}\left( t\right) +\alpha .
\end{equation}
We consider the stochastic dynamics generated by the formal Hamiltonian

\begin{equation}
\Upsilon _{t}=Ca_{t}^{+}+C^{\dagger }a_{t}^{-}+F
\end{equation}
where $C$ and self-adjoint $F$ are operators on $\mathcal{H}_{0}$. We are
required to ``normal order'' the unitary process $U_{t}=\mathbf{\vec{T}}\exp
\left\{ -i\int_{0}^{t}ds\,\Upsilon _{s}\right\} $ for the given state and
this means normal order the $a_{1}^{\pm }\left( t\right) $ and the $%
a_{2}^{\pm }\left( t\right) $. By using the same technique as in (13) we
find that, for instance, $\left[ a_{1}^{-}\left( t\right) ,U_{t}\right]
=-i\int_{0}^{t}ds\,\left[ a_{1}^{-}\left( t\right) ,\Upsilon _{s}\right]
U_{s}=-i\kappa xCU_{t}$. We can deduce the rewriting rules 
\begin{eqnarray*}
a_{1}^{-}\left( t\right) U_{t} &=&U_{t}\left( a_{1}^{-}\left( t\right)
-i\kappa xC\right) ; \\
a_{2}^{-}\left( t\right) U_{t} &=&U_{t}\left( a_{2}^{-}\left( t\right)
-i\kappa z^{\ast }C-i\kappa yC^{\dagger }\right) .
\end{eqnarray*}
The qsde $\frac{dU_{t}}{dt}=-i\Upsilon _{t}U_{t}$ can be then be rewritten
as 
\begin{eqnarray*}
\frac{dU_{t}}{dt} &=&\,-i:\Upsilon _{t}U_{t}:\,-iCyU_{t}\left( -i\kappa
z^{\ast }C-i\kappa yC^{\dagger }\right) \\
&&-iCxU_{t}\left( -i\kappa xC\right) -iC^{\dagger }zU_{t}\left( -i\kappa
z^{\ast }C-i\kappa yC^{\dagger }\right)
\end{eqnarray*}
where :$\Upsilon _{t}U_{t}:$ is the reordering of $\Upsilon _{t}U_{t}$
placing all creators $a_{1}^{+}\left( t\right) $ and $a_{2}^{+}\left(
t\right) $ to the left and all annihilators $a_{1}^{-}\left( t\right) $ and $%
a_{2}^{-}\left( t\right) $ to the right. Rearranging gives 
\begin{eqnarray}
\frac{dU_{t}}{dt} &=&\,-iC\left( xa_{1}^{+}\left( t\right)
U_{t}+yU_{t}a_{2}^{-}\left( t\right) +z^{\ast }a_{2}^{+}\left( t\right)
U_{t}+\alpha ^{\ast }U_{t}\right)  \notag \\
&&-iC^{\dagger }\left( xU_{t}a_{1}^{-}\left( t\right) +ya_{2}^{+}\left(
t\right) U_{t}+zU_{t}a_{2}^{-}\left( t\right) +\alpha U_{t}\right)  \notag \\
&&-\left[ iF+\kappa \left( \left( n+1\right) C^{\dagger }C+nCC^{\dagger
}+m^{\ast }CC+mC^{\dagger }C^{\dagger }\right) \right] U_{t}.
\end{eqnarray}
To obtain an equivalent Hudson-Parthasarathy qsde, we introduce quantum
Brownian motions $A_{t}=\int_{0}^{t}\left( xa_{1}^{-}\left( s\right)
+ya_{2}^{+}\left( s\right) +za_{2}^{-}\left( s\right) \right) ds$ with the
under standing that, for $R_{t}$ adapted, $R_{t}\otimes dA_{t}=\left(
xR_{t}a_{1}^{-}\left( t\right) +ya_{2}^{+}\left( t\right)
R_{t}+zR_{t}a_{2}^{-}\left( t\right) \right) $ and $R_{t}\otimes
dA_{t}^{\dagger }=\left( xa_{1}^{+}\left( t\right)
R_{t}+yR_{t}a_{2}^{-}\left( t\right) +z^{\ast }a_{2}^{+}\left( t\right)
R_{t}\right) $. Then 
\begin{equation}
dU_{t}\equiv -iCU_{t}\otimes dA_{t}^{\dagger }-iC^{\dagger }U_{t}\otimes
dA_{t}-GU_{t}\otimes dt
\end{equation}
where $G=i\left( F+\alpha ^{\ast }C+\alpha C^{\dagger }\right) +\kappa
\left( \left( n+1\right) C^{\dagger }C+nCC^{\dagger }+m^{\ast
}CC+mC^{\dagger }C^{\dagger }\right) $. Note that the quantum It\^{o} table
will be 
\begin{eqnarray}
dA_{t}dA_{t}^{\dagger } &=&\gamma \left( n+1\right) \,dt;\qquad
dA_{t}^{\dagger }dA_{t}=\gamma n\,dt;  \notag \\
dA_{t}^{\dagger }dA_{t}^{\dagger } &=&\gamma m^{\ast }\,dt;\qquad
dA_{t}dA_{t}=\gamma m\,dt.
\end{eqnarray}
It is readily shown, either by normal ordering or by means of the quantum
stochastic calculus, that the stochastic Heisenberg equation is then 
\begin{equation}
dJ_{t}\left( X\right) =-iJ_{t}\left( \left[ X,C^{\dagger }\right] \right)
\otimes dA_{t}^{\dagger }-iJ_{t}\left( \left[ X,C\right] \right) \otimes
dA_{t}+J_{t}\left( L\left( X\right) \right) \otimes dt
\end{equation}
where 
\begin{equation}
L\left( X\right) =\gamma \left\{ \left( n+1\right) C^{\dagger
}XC+nCXC^{\dagger }+mCXC+m^{\ast }C^{\dagger }XC^{\dagger }\right\}
-XG-G^{\dagger }X.
\end{equation}

Finally the master equation is obtained by duality: $\frac{d}{dt}\varrho
=L^{\prime }\left( \varrho \right) $ where $tr\left\{ \varrho L\left(
X\right) \right\} \equiv tr\left\{ L^{\prime }\left( \varrho \right)
X\right\} $.

\end{document}